\documentclass[sigconf]{acmart}

\usepackage{framed}
\usepackage[strict]{changepage}
\usepackage{subcaption}
\usepackage{placeins}
\usepackage{lettrine} 
\definecolor{formalshade}{rgb}{0.95,0.95,1}
\definecolor{darkblue}{rgb}{0.36,0.54,0.66}

\newenvironment{formal}{%
  \MakeFramed{\advance\hsize-\width\FrameRestore}%
  \noindent\hspace{-4.55pt}
  \begin{adjustwidth}{}{7pt}%
  \vspace{2pt}\vspace{2pt}%
}
{%
  \vspace{2pt}\end{adjustwidth}\endMakeFramed%
}

\AtBeginDocument{%
  }

\copyrightyear{2026}
\acmYear{2026}
\setcopyright{rightsretained}
\acmConference[The 3rd InterAI Workshop at CHI 2026]{The 3rd InterAI Workshop: "Interactive AI for Human-Centered Robotics at ACM CHI 2026}{April 13, 2026}{Barcelona, Spain}
\acmBooktitle{The 3rd InterAI Workshop: "Interactive AI for Human-Centered Robotics at ACM CHI 2026, April 13--17, 2026, Barcelona, Spain}

\acmISBN{978-1-4503-XXXX-X/2018/06}

\usepackage{subcaption}
\usepackage{placeins}
\begin{document}

\title[Directing the Robot]{Directing the Robot: Scaffolding Creative Human–AI–Robot Interaction}


\author{Jordan Aiko {Deja}} 
\orcid{0000-0001-9341-6088}
 \affiliation{%
  \institution{De La Salle University}
  \city{Manila}
  \country{Philippines}
  }
\email{jordan.deja@dlsu.edu.ph}

\author{Isidro {Butaslac} III} 
\orcid{0000-0003-1172-7611}
 \affiliation{%
  \institution{Nara Institute of Science and Technology}
  \city{Ikoma}
  \country{Japan}
  }
\email{isidro.b@naist.ac.jp}

\author{Nicko Reginio {Caluya}} 
\orcid{0000-0003-4924-958X}
 \affiliation{%
  \institution{Ritsumeikan University}
  \city{Osaka}
  \country{Japan}
  }
\email{nicko@fc.ritsumei.ac.jp}

\author{Maheshya Weerasinghe}
\orcid{0000-0003-2691-601X}
\affiliation{%
  \institution{University of Primorska}
  \city{Koper}
  \country{Slovenia}
 \postcode{6000}}
\email{maheshya.weerasinghe@famnit.upr.si}


\begin{abstract}

Robots are moving beyond industrial settings into creative, educational, and public environments where interaction is open-ended and improvisational. Yet much of human–AI–robot interaction remains framed around performance and efficiency, positioning humans as supervisors rather than collaborators. We propose a re-framing of AI interaction with robots as scaffolding: infrastructure that enables humans to shape robotic behaviour over time while remaining meaningfully in control. Through scenarios from creative practice, learning-by-teaching, and embodied interaction, we illustrate how humans can act as executive directors, defining intent and steering revisions, while AI mediates between human expression and robotic execution. We outline design and evaluation implications that foreground creativity, agency, and flow. Finally, we discuss open challenges in social, scalable, and mission-critical contexts. We invite the community to rethink interacting with Robots and AI not as autonomy, but as sustained support for human creativity.
\end{abstract}


\begin{teaserfigure}
\centering
  \includegraphics[width=0.95\textwidth]{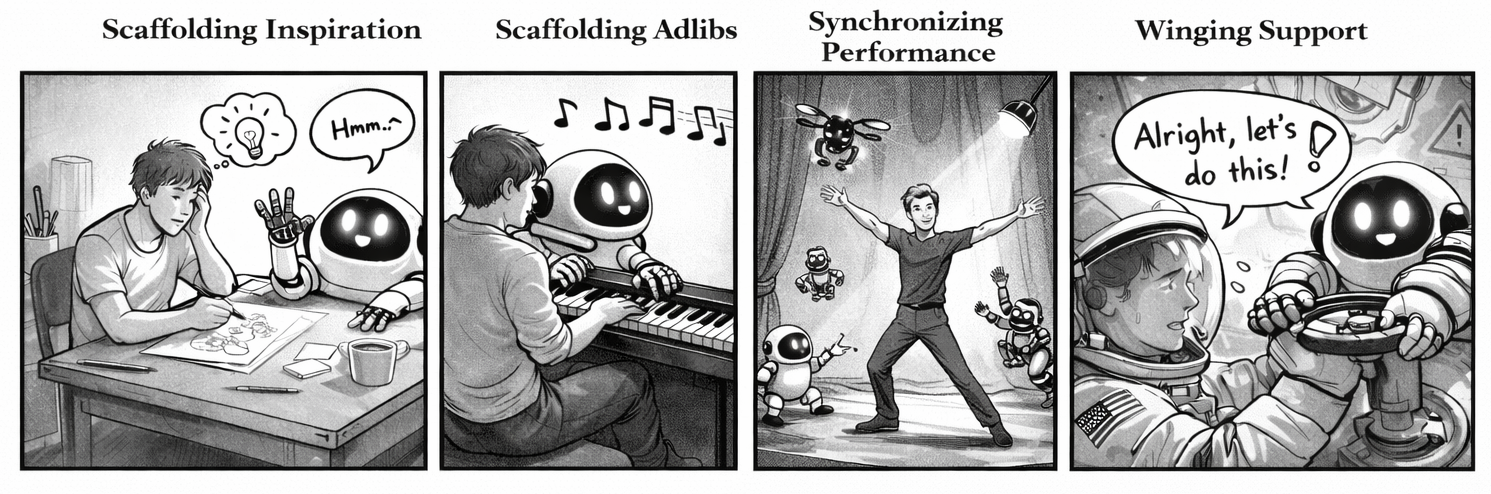}
 \caption{Illustrative scenarios of scaffolding in creative human–AI–robot interaction. From left to right: (1) Scaffolding Inspiration (robot gestures support ideation); (2) Scaffolding Adlibs (robot improvises in musical performance); (3) Synchronizing Performance (ground and aerial robots in stage collaboration); and (4) Winging Support (robot assistance under time pressure). Across scenarios, AI scaffolds creativity and action while preserving human agency. \textit{Illustration generated using generative AI and refined by the authors}.}
  \label{fig:teaser}
\end{teaserfigure}



\maketitle

\section{Introduction and Background}

\par Robots are increasingly becoming part of daily life, expanding beyond industrial settings~\cite{hentout2019human} into public spaces, educational environments~\cite{pasalidou2025augmented}, and creative contexts~\cite{robinson2023robotic}. As robots move closer to people in both physical and social settings, interaction with them is no longer exceptional but inevitable. Advances in Artificial Intelligence (AI) have been central to this shift, often by making human--robot interaction more accessible through the automation of behaviours, optimisation of performance, and reduction of human effort in repetitive or routine tasks. While effective, these approaches have also shaped dominant design paradigms that prioritise efficiency and autonomy over richness of intent.

\par However, framing AI primarily as a tool for automation and optimisation limits how humans and robots can collaborate in open-ended settings. In such framings, humans are often positioned as supervisors or end-users, while decision-making authority and interpretive control are increasingly delegated to the system. In this context, interpretative control is defined as how the human decides the execution similar to how the director of photography decides on the color toning and angles when shooting a film. This diminishes the value of human autonomy ~\cite{formosa2021autonomy}, and limits active engagement, exploration, and creative collaboration, particularly in expressive, educational, or improvisational contexts. 

\par Prior work has explored improvisation, mixed-initiative interaction, and co-creative human--AI systems~\cite{hoffman2019evaluating, horvitz1999principles, davis2013human, argall2009survey}. However, scaffolding in these systems is often treated as either a technical mechanism for learning or a temporary design strategy used to compensate for system limitations. In contrast, we argue for treating scaffolding as a ``core interaction principle'' in human--AI--robot collaboration; one that explicitly structures how agency, interpretation, and control are distributed over time. In this case, we look at not leaving it all to the AI or robot, but rather having the original idea come from the human and both human and AI collectively build on their ideas. 

\par Building on this perspective, we propose a re-framing in which interactive AI scaffolds collaboration by positioning humans as ``executive directors'' of robotic behaviour. In this role, humans define intent, make interpretive judgments, and steer iterative revisions, while interactive AI mediates between human expression (e.g., language, gesture, demonstration) and robotic execution (e.g., motion, coordination, embodied action).  Rather than handing control away from the human, AI supports thinking, decision-making, and creative flow, while robots act as expressive and physical performers of human-defined intent. This perspective foregrounds human agency and re-frames interaction as an ongoing creative process rather than a supervisory task. Specifically, this paper contributes (1) a re-framing of interactive AI as scaffolding for creative human-robot interaction, (2) illustrative scenarios spanning creative and mission-critical contexts (as shown in ~\autoref{fig:teaser}), and (3) design and evaluation implications centered on agency and flow.

\section{Rethinking Design and Evaluation Through Use Cases}

\par To ground the idea of scaffolding as an interaction principle, we consider use cases in which creativity, exploration, and improvisation are central. In many such contexts, robots are not continuously autonomous actors, but remain available as expressive resources that can be activated, redirected, or reinterpreted by humans as needed. Interactive AI mediates this availability, allowing humans to focus on intent, timing, and meaning rather than low-level control. In this sense, interaction unfolds as a process of direction rather than execution. Across these scenarios, humans act as executive directors: they articulate high-level goals, shape interpretation, and intervene iteratively as outcomes emerge. Interactive AI supports this role by translating underspecified human input into executable robotic behaviour, managing ambiguity and coordination while leaving authorship and decision-making with the human.

\begin{formal}
\textbf{Thought 1: Preserve interpretive control.} 
Interactive AI must support human intent without displacing authorship or the ability to redirect action.
\end{formal}
\par One illustrative case involves teaching or shaping robot motion through few-shot demonstrations~\cite{argall2009survey}. A human may imagine a sequence of movements or expressive qualities they wish a robot to perform, but may be constrained by physical ability, scale, or precision. Rather than requiring exact replication, interactive AI can scaffold this process by interpreting partial demonstrations, smoothing transitions, and generating feasible motion variants that preserve the intended structure and emphasis of the human (~\autoref{fig:teaser} Frame 1). Here, the human remains responsible for defining what the movement should convey, while AI manages how that intent is realized in robotic execution.


\par Also as in emerging learning-oriented contexts, where humans do not merely demonstrate tasks, but iteratively shape how robots collaborate with them over time. For instance, in collaborative assembly settings, researchers have developed a ``Teaching-Learning-Collaboration (TLC)'' model in which a human demonstrates a task and communicates high-level intent and preferences to a robot collaborator~\cite{wang2018facilitating}. The robot uses this information to build task knowledge via learning from demonstrations and then actively assists the human in completing shared work. 
Such models illustrate how humans can act as executive directors: guiding, teaching, and co-revising robotic behaviour while interactive learning enables robots to translate human intent into embodied collaborative action without displacing control.

\par These examples suggest several design implications for interactive AI as scaffolding. First, systems should prioritise ``responsiveness over correctness'', enabling interaction to continue even when human input is partial, ambiguous, or evolving. Second, interfaces that leverage familiar forms of expression such as language, gesture, or embodied demonstration, can reduce cognitive overhead by allowing humans to direct robots in ways that align with how they already communicate intent. In such cases, AI acts less as an autonomous decision-maker and more as a translator between human expression and robotic performance.

\begin{formal}
\textbf{Thought 2: Prioritise responsiveness over optimisation.} 
Interactive AI should sustain creative flow under evolving input rather than converging on a fixed optimal solution.
\end{formal}

\par This framing is particularly relevant in domains such as human--drone and multi-robot interaction, where direct manual control is often impractical. In performative or large-scale settings, such as stage productions involving drones or zoomorphic robots~\cite{madill2025playing, mckendrick2023waiting}, humans may cue, redirect, or improvise in real time using gestures or brief verbal instructions. Interactive AI can scaffold coordination and timing across multiple robots (~\autoref{fig:teaser} Frame 3), enabling collective behaviour that remains responsive to human direction rather than pre-scripted autonomy. Shape, motion, and embodiment further support imagination and storytelling, allowing robots to function as expressive performers rather than independent agents~\cite{jeung2025shape, wang2025puppetline}.


\par Rethinking design this way also requires rethinking evaluation. Traditional HRI metrics such as efficiency, error rate, or task completion time are poorly suited to interactions centered on exploration and improvisation. Instead, evaluation should ask whether systems support humans in experiencing themselves as directing the interaction--able to reinterpret, intervene, and steer outcomes as they unfold. Metrics that capture flow~\cite{csikszentmihalyi1990flow}, perceived agency, and sustained engagement, such as the Creativity Support Index~\cite{cherry2014quantifying}, provide more appropriate lenses for assessing whether interactive AI functions effectively as scaffolding rather than as automation.

\begin{formal}
\textbf{Thought 3: Measure agency as a first-class outcome.} 
Evaluate scaffolding by users’ sense of authorship and flow, not just efficiency.
\end{formal}



\par Framing evaluation around creativity and agency also serves a broader purpose. By explicitly valuing human direction and interpretive control, interactive AI systems resist narratives of robots replacing or overpowering human collaborators. We can do this by designing systems that support humans in articulating intent, maintaining creative flow, and remaining accountable for outcomes sometimes through something as simple as a gesture, a cue, or a brief corrective intervention~\cite{sta2025set}. In this way, evaluation becomes a design commitment rather than a post-hoc measurement.


\section{Open Challenges and Broader Implications}

\par We also raise a set of open challenges that extend beyond individual use cases. 
One challenge concerns social interaction and physical proximity in close human--robot collaboration~\cite{sawabe2022robot}. When humans direct robots through naturalistic cues such as gesture, touch, or expressive movement, the boundaries of control and consent may become ambiguous. In such interactions, responsibility does not lie solely in issuing commands, but in shaping interpretation and responding to system behaviour as it unfolds. Designing interactive AI that can scaffold expressive direction while making the boundaries of agency legible remains an open problem. 

\par Another challenge arises when scaling the direction from a single robot to multiple robots or swarms. As humans gesture, speak, or demonstrate intent to multiple robots simultaneously, how should that intent be preserved, transformed, or distributed throughout the system~\cite{graf2024distributed, walker2018communicating}? Although interactive AI can mediate coordination and execution, increased scale also risks diluting human interpretive control. This raises questions about how much autonomy can be distributed without eroding the human’s role as director, particularly in domains where improvisation and responsiveness are critical.

\par These challenges are further amplified in emerging creative domains where robots function not only as interactive agents, but also as active materials for design and expression. In contexts involving non-rigid or deformable robotic systems~\cite{boem2019non}, deformable robotic systems~\cite{suzuki2022augmented}, or expressive robotic bodies, interaction unfolds through ongoing negotiation rather than discrete instruction. 
Similar dynamics arise in creative practices such as robotic dance, performance, or musical improvisation~\cite{hoffman2010gesture}; for example, when a robot participates in piano improvisation alongside a human performer (similar to that in ~\cite{deja2022vision} (~\autoref{fig:teaser} Frame 2)). 
Interactive AI must therefore scaffold exploration and interpretation without prematurely stabilising behaviour or constraining expressive possibility.


\par While we ground our discussion in creative contexts, the scaffolding lens extends to any domain where intent evolves over time. Mission-critical contexts introduce additional tension between direction and safety. In time-pressured situations such as disaster response or emergency coordination, humans may need to improvise or ``wing it'' under uncertainty~\cite{jetter2025mixed}. Interactive AI can scaffold rapid interpretation and execution, but over-directing or delaying action risks undermining human intuition and accountability (~\autoref{fig:teaser} Frame 4). Conversely, insufficient mediation may expose users to safety or trust failures~\cite{grzeszczuk2025building}. Designing systems that support human-directed improvisation, so they can ``wing-it like McGyver'' during such scenarios, while maintaining transparency, trust, and responsibility remains a central challenge.

\par 
Taken together, if humans are to remain executive directors rather than supervisors of autonomous systems, then interactive AI must be designed to support interpretation, intervention, and accountability in these systems. Addressing these challenges will require rethinking not only system capabilities, but also how agency and responsibility are made visible and negotiable in human--AI--robot collaboration.

\section{Conclusion}

\par As interactive AI and robotics mature, the central question for human--robot interaction is no longer how to maximise autonomy, but how to design systems that know when not to take over. In creative, learning-oriented, and time-critical contexts, effective collaboration is less dependent on optimisation than on the ability to improvise, reinterpret, and adapt as situations unfold. We have argued for re-framing interactive AI as scaffolding: infrastructure that enables humans to shape robotic behaviour while remaining meaningfully in control. By positioning humans as executive directors and AI as mediating partners, this perspective foregrounds interpretation, intervention, and accountability as first-class interactional concerns. Rather than prescribing a specific architecture, we offer a conceptual lens for reasoning about initiative and control in human--AI--robot collaboration. We invite the community to rethink automation-centric narratives and explore how interactive AI can function as enduring scaffolding for human creativity, learning, and responsible action with robots.

\bibliographystyle{ACM-Reference-Format}
\bibliography{references}

\appendix









\end{document}